\documentclass[a4paper,twocolumn]{esapub2005}
\usepackage{epsfig}
\usepackage{graphics}

\begin{document}

\title{The seismology programme}
\author[1]{E. Michel} 
\author[1]{A. Baglin}  
\author[1]{M. Auvergne}
\author[1]{C. Catala}
\author[2]{C. Aerts}
\author[3]{G. Alecian} 
\author[4]{P. Amado}
\author[5]{T. Appourchaux} 
\author[2]{M. Ausseloos}
\author[6]{J. Ballot} 
\author[1]{C. Barban} 
\author[5]{F. Baudin} 
\author[7]{G. Berthomieu} 
\author[5]{P. Boumier} 
\author[8]{T. B\"ohm}
\author[2]{M. Briquet}
\author[8]{S. Charpinet} 
\author[9]{M. S. Cunha}
\author[10]{P. De Cat} 
\author[1]{M.A. Dupret} 
\author[11]{J. Fabregat}
\author[12]{M. Floquet}
\author[10]{Y. Fr\'emat}
\author[4]{R. Garrido}
\author[6]{R. A. Garcia}
\author[1]{M.-J. Goupil} 
\author[13]{G. Handler}
\author[12]{A.-M. Hubert}
\author[14]{E. Janot-Pacheco}
\author[6]{P. Lambert}
\author[12]{Y. Lebreton}
\author[8]{F. Ligni\`eres}
\author[1]{J. Lochard}
\author[15]{S. Mart\'in-Ruiz} 
\author[16]{P. Mathias}
\author[1]{A. Mazumdar}
\author[13]{P. Mittermayer}
\author[17]{J. Montalb\'an}
\author[9]{M. Monteiro}
\author[7]{P. Morel}
\author[1]{B. Mosser}
\author[1,4]{A. Moya}
\author[12]{C. Neiner}
\author[6]{P. Nghiem}
\author[17]{A. Noels}
\author[13]{J. Oehlinger}
\author[15]{ E. Poretti}
\author[7]{J. Provost} 
\author[18]{J. Renan de Medeiros} 
\author[2]{J. de Ridder}
\author[8]{M. Rieutord}
\author[19]{T. Roca-Cort\'es}
\author[20]{I. Roxburgh}
\author[1]{R. Samadi}
\author[17]{R. Scuflaire}
\author[4,1]{J.C. Suares}
\author[17]{S. Th\'eado}
\author[17]{A. Thoul}
\author[7]{T. Toutain}
\author[6]{S. Turck-Chi\`eze}
\author[2]{K. Uytterhoeven}
\author[8]{G. Vauclair}
\author[8]{S. Vauclair}
\author[13]{ W.W. Weiss}
\author[13]{K. Zwintz}
\author{and the COROT Team}

\affil[1]{ 
Laboratoire d'Etudes Spatiales et d'Instrumentation pour l'Astrophysique, UMR~8109, Obs. de
Paris, Meudon, France
}
\affil[2]{ 
Instituut of Astronomy, voor Sterrenkunde Katholieke Universiteit Leuven,  Belgium 
}
\affil[3]{ 
Laboratoire Univers et Th\'eories, UMR~8102, Observatoire de
Paris, 92190, Meudon, France
}
\affil[4]{ 
Instituto de Astrof\'isica de Andaluc\'ia (CSIC), Granada, Spain 
}
\affil[5]{ 
Institut d'Astrophysique Spatiale, CNRS/Univ. Paris XI, UMR~8617, Orsay, France 
}
\affil[6]{ 
CEA/DAPNIA: Service d'Astrophysique, UMR~7158, Gif-sur-Yvette, France
}
\affil[7]{ 
D\'epartement Cassiop\'ee, UMR~6202, Observatoire de la C\^ote d'Azur, Nice, France
}
\affil[8]{ 
Laboratoire d'Astrophysique de l'Observatoire Midi-Pyr\'en\'ees, Univ. Paul-Sabatier, UMR~5572, Toulouse, France
}
\affil[9]{ 
Centro de Astrof\'isica da Univ. de Porto, Porto, Portugal
}
\affil[10]{ 
Royal Observatory of Belgium, Brussels, Belgium
}
\affil[11]{ 
Univ. de Valenc\'ia, Instituto de Ciencia de los Materoales, Valenc\'ia, Spain
}
\affil[12]{ 
GEPI, UMR~8111, Observatoire de Paris, Meudon, France
}
\affil[13]{ 
Institute f\"{u}r Astronomie, Univ. of Vienna, Wien, Austria
}
\affil[14]{ 
Instituto de Astronomia, Geof\'isica e Ci\^encias Atmosf\'ericas, Univ. de S\~ao Paulo, S\~ao Paulo, Brazil
}
\affil[15]{ 
INAF-Osservatorio Astronomico di Brera, Merate, Italy
}
\affil[16]{ 
D\'epartement Gemini, UMR~6203, Observatoire de la C\^ote d'Azur, Nice, France
}
\affil[17]{ 
Institut d'Astrophysique et G\'eophysique, Univ. de Li\`ege, Li\`ege, Belgium
}
\affil[18]{ 
Departamento de Fisica, Univ. Federal do Rio Grande do Norte, Natal, Brazil
}
\affil[19]{ 
Instituto de Astof\'isica de Canarias, Univ. de la Laguna, La Laguna, Tenerife, Spain
}
\affil[20]{ 
Astronomy Unit, Queen Mary, Univ. of London, London, United Kingdom
}

\maketitle

\begin{abstract}

We introduce the main lines and specificities of the CoRoT Seismology Core Programme.  
The development and consolidation of this programme has been made in the framework of the
CoRoT Seismology Working Group. With a few illustrative examples, we show how CoRoT data will help
to address various problems associated with present open questions of stellar structure and evolution.

\end{abstract}

\section{Introduction: the Seismology programme and the CoRoT SWG}.

The main lines of the CoRoT seismology programme have been designed very early (Catala et al. 1995,
Baglin et al. 1998) with
strong specificities, in terms of requirements on the precision of frequency measurements, duration of the runs.
Then, 
in order to develop this programme and optimize the scientific return,
the CoRoT Seismology Working Group has been settled 
at the CoRoT kick-off meeting (1998, in Nice). 
The CoRoT SWG is intended to supply the required expertise and promote the investigation
of relevant 'hard points'. This work is at the base of the mission profile determination and of the targets selection.
The SWG counts approximately 90 members in 20 institutes. The activity of the SWG has been regularly adapted to the needs associated 
to the evolution of the CoRoT project. 

In Sect.2, we sketch out the scientifc context for a project of stellar seismology like CoRoT. 
First (Sect.2.1), we try to give a flavour of
the various scientific questions at stake. Then (Sect.2.2), we briefly come back on what has already been achieved so far and
how this experience has logically lead the stellar international community to plan dedicated observations
from space. 

The CoRoT Seismology Core programme is described in Sect.3. For both Solar-like pulsators and ''classical'' pulsators
(resp. Sect.3.1 and Sect.3.2),
we present a few results obtained in the framework of the preparation of the CoRoT Seismology Programme. These results are
selected to illustrate the preparatory work which has been made to investigate
the various aspects of the Seismology Programme. The list is not pretended to be exhaustive and this is not our purpose either
to develop these points in details.  Some of them are developed further in this volume.

Finally, in Sect.3.3, we come back on a few aspects of the fields and targets selection process. 

\section{The context}

\subsection{Open questions of stellar physics and the  
seismology promiss}

Stars are one of the main constituents of the Universe; they are also one of the major sources of information
about it and thus an unavoidable subject of study. 
Nearly every field of astrophysics uses results of stellar structure and evolution theory, to estimate, for example,
the age of globular clusters which give an essential piece of information on the age of the Universe, or to understand
the origine of the chemical elements or the history of the Sun and of the solar system.

The main lines of stellar structure and evolution have been understood by
confrontation of observables coming from the surface
of stars and theoretical modelling calling to a wide panel of various fields of physics.
Our understanding of stellar evolution and our capability to describe it precisely is
thus suffering large uncertainties, due to the fact that a star is a complex object, involving
a large number of physical processes still poorly understood.

\begin{figure}[ht]
   \begin{center}
    \includegraphics[angle=-90,width=0.9\hsize]{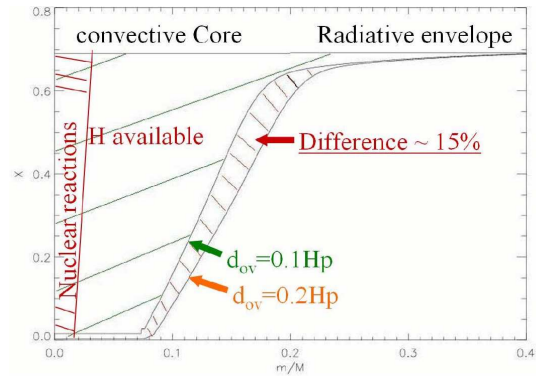}
   \end{center}
\caption{ Hydrogen mass fraction profile (X) for a 1.8M$_{\odot}$ stellar model at the end of
Main Sequence evolution, for two illustrative values of the overshooting parameter d${_{ov}}$. Hached areas
are proportional to the amount of hydrogen available for nuclear reactions on the Main Sequence,  
showing a difference
of $15\%$ between the two cases considered here.   
\label{xandov}}
\end{figure}

Considering the upper part of the main sequence for instance, which is characterized by the existence
of a convective core, one of the most debated open question is whether and how 
the central region mixed by convection is extended 
by the so-called overshooting process. This point
alone, by the change induced in the amount of hydrogen available for central nuclear reactions (see
Fig.~\ref{xandov}),
 is responsible for an uncertainty
which can reach $30$ to $50\%$ in the age estimate of all stars with mass higher than
 $\sim$ 1.1M$_{\odot}$ (for solar composition).

For the lower part of the main sequence ( M $\leq$ 1.4M${\odot}$), characterized by an extended outer
convective zone below the surface, one of the most prominent open questions deals with the
efficiency of the heat transport in the upper convective regions. There, the density is so low
that the heat transport efficiency depends severely on the description of the convective transport
process.
Since a fully consistent description of this process is still out of reach, 
the uncertainty in determining
the temperature gradient becomes important and hampers severely our description of stars.

In spite of several tentative refinements, these two processes are widely considered at the moment
in the modeling by simple one-parameter crude descriptions.

Beyond these two points, segregation of the different
chemical species induced by gravitational and radiative forces are considered as a non negligible factor
in several classes of objects, with a prominent manifestation in the
surface anomalies observed in A stars. Impressive efforts have been made to implement these aspects
(e.g. Michaud 2004), but they have been only confronted so far to surface classical observables and would benefit 
additional observational constraints.

Mass loss, meridional circulation and turbulence, their influence on angular momentum
evolution, their interaction with previously quoted diffusion of chemical elements
constitute another active front for research in this field (e.g. Talon 2004, Vauclair and Th\'eado 2003,
Th\'eado and Vauclair 2003),
but here again, classical observables hardly can constrain such processes.
On top of this, one can consider the effect of magnetic field and its potential interaction
with the previous mechanisms (Alecian 2004, Mathis \& Zahn 2005).

In the pre-Main Sequence phase, it is still an open question to know how the angular 
momentum evolves and what is its influence on the structure at the beginning of the Main Sequence phase.

At the other end of stellar evolution, the chemical composition profile of
the strongly stratified structure of white dwarfs is holding
the signature of e.g.
$C^{12}-O^{16}$ poorly known reaction rate or mixing processes at work during the red giant phase.

In all these cases, stellar oscillations are expected to bring relevant additional constraints. 
Oscillations have been observed in stars representative of approximately all  mass ranges and
evolution stages, from the PMS stage, to the white dwarf cooling sequence,
including Main Sequence, horizontal branch and red giant phases.
New classes of stellar pulsators are still regularly discovered,
and pulsation looks now more like the rule rather than the exception.
This definitely suggests that seismology is a promising tool to improve our understanding of the various
physical processes at work in stars.

Stars are generally seen pulsating on a more or less extended set of eigenmodes.
These eigenmodes are standing waves established inside
the stars by essentially two types of propagation waves: pressure waves, for which the restoring force
is dominantly the pressure gradient, and gravity waves, for which the restoring force is essentially the
buoyancy.

\begin{figure}[ht]
   \begin{center}
    \includegraphics[angle=-90,width=0.9\hsize]{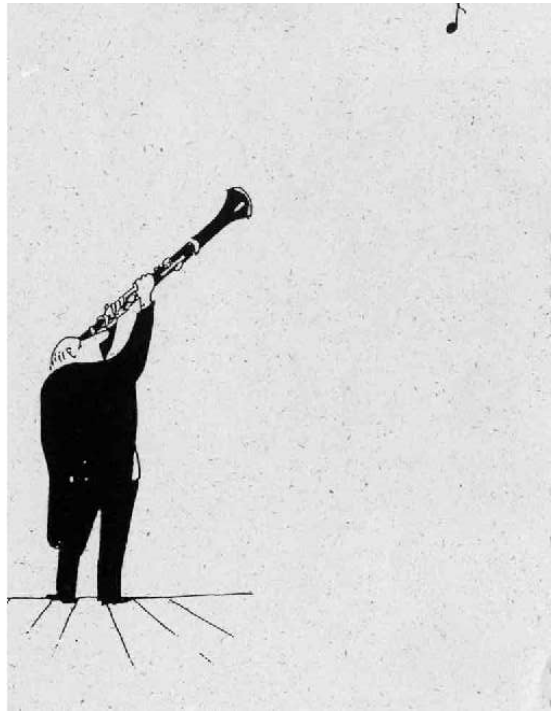}
   \end{center}
\caption{ 
'Et pourtant, elles pulsent!', Philippe Delache 1992.
\label{phil}}
\end{figure}

The  associated eigenfunctions can generally be decomposed in a product of a function $f(r)$ describing
the radial dependence and the angular dependence expressed
in terms of spherical harmonics (see e.g. Unno et al. 89).
Each eigenmode can thus be characterized by three integers: $l$ and $m$, the degree and the azimuthal order
of the spherical harmonics, and the radial order $n$, a count of the nodes of function $f(r)$.

The basic idea of stellar seismology is to use these new observables (frequencies, amplitudes,
profiles,phase, ...) which are sensitive to the interior structure in a differential way,
to help constraining the stellar structure and its evolution.

\subsection {From space and from the ground, state of the art of seismology}

  The case of our closest star is
illustrative of the technique, though it has its
own specificity.
Observations of the Sun, from space and from the ground, have brought
eigenfrequencies determined with precision up to a few
nHz, thanks to very long and dense 
observation sequences (longer than 10 years with a 93\% duty for SOHO,
20 years with Iris, 30 years with 
Bison, see e.g. Salabert et al. 2004; Garcia et al. 2004, 2005).
The high spatial resolution allows to
detect millions of eigenmodes.
The results are impressive. The measurement of
the rotation of the Sun with respect to radius
and latitude down to the inner 20\% central
region (see e.g. Couvidat et al. 2003) has confirmed the
existence of the tachocline, but has
also assessed the solid rotation regime inside
the radiative zone, which resists interpretation
in terms of angular momentum transfer mechanism.
Solar seismology has shown its ability to solve
physical controversies in measuring precisely the
central temperature, constraining
to accept a neutrino mass (see e.g. Turck-Chi\`eze et al. 2001, 2004). 

Great efforts are
developed to track gravity modes (e.g.
Picard (see Thuillier et al. 2006), Golf-NG, see e.g. Turck-Chi\`eze et al. 2005,
2006,
see also Appourchaux et al. 2000)
which would allow to probe the
not yet resolved very inner
part of the Sun.

Generalizing seismology techniques to all kind
of stars has been a leading idea for stellar
astronomers.Though the accuracy and resolution
obtainted on the Sun look out of reach for
distant stars, the large number of objects will
allow a global vision of stellar evolution.
Let us mention briefly the main development lines
in this domain.

%\subsection{the extension to other solar-like pulsators}

Solar-like oscillations are expected to exist in
a large number of stars, in fact all F and early
G stars, with amplitudes equal to or
larger than in the Sun ( see e.g. Houdek et al. 1999). 
These amplitudes remain
intrinsically small (a few ppm in photometry and
a few tens of cm/s in radial velocity)
and the search for such oscillations in stars
other than the Sun has been a long quest for the
international community.
After the first well-established detection about
7 years ago (Martic et al. 1999, Barban et al. 1999),
oscillations have
been found in a few more stars (see Bedding and Kjeldsen 2006
and references therein), thanks in
particular to new spectrographs on large
telescope like HARPS (see e.g. Mosser et al. 2004,
Bouchy et al. 2005, Santos et al. 2004).
These observations have provided tests for the
theory of excitation of the modes in comparing
predicted and observed amplitudes (Samadi et al.
2003 and ref therein). Interpretation of
individual frequencies remains risky, due to the
small signal to noise ratio and to the short
duration of the observing runs. 
In the  very
favorable case of Alpha Cen A and B,
for which oscillations have been found in both
objects, several modelling studies  have been made
(Thevenin et al. 2002, Eggenberger et al. 2004,
Miglio and Montalb\'an 2005),
using seismic information and global parameters
obtained from binarity
characterization,
and interferometry measurements.
Miglio and Montalb\'an (2005), discuss 
the interest and the impact of the different observables
on the study. In the actual state of the art however,
it hardly leads to a firm conclusion in terms of physics,
partly because, even in this case, most of the observable 
can still be questioned
to some extent,
but also because our experience in this domain still has to be built
and secured by confrontation with better data for a larger set of objects.

%\subsection{Classical "large amplitude" variables}

For a large number of classical variable stars
($\delta$ Scuti, white dwarfs, sdB, Be, $\gamma$
Dor,...) some modes show amplitudes
larger than $\sim 10^{-3}$ in photometry ($\sim$ 100m/s in
radial velocity) and can be observed from the
ground.
Observations are planned regularly, often within
coordinated multisite campaigns (STEPHI,
WET, DSN,... see e.g. Li et al. 2004, Breger et al 2005, Vauclair et al. 2003,
Mathias et al. 2004,...),
which allow to reach
these detection levels and to obtain satisfying
time coverage and associated window function.

Interpretation in terms of modelling internal
structure reaches different stages for different
types of objects.
For white dwarfs, 
various studies tend to
determine the stellar parameters: total mass, mass of the envelope,
rotation periode (see e.g. Pech et al. 2006) and
the characteristic time scales on the
cooling sequence.
For $\delta$ Scuti and Be stars,  fast rotation,
very common in these classes of objects,
has to be taken into account under several
aspects of the modelling. This has driven
theoretical developments, numerical
implementation, and is now a crucial aspect of
most of the present studies (Soufi et al. 1998,
Ligni\`eres et al. 2003,
Ligni\`eres et al. 2006, 
Reese et al. 2006).

For $\gamma$ Dor objects, (which were not known
10 years ago), a coherent picture for the driving source
of the oscillations has been reached only recently
(Dupret et al. 2005).
For all these objects however, it is fair to say that seismic
interpretation has not yet reached the 'exploitation' level
in terms of scientific return on physical processes at stake.

Space very early appeared as a predilection place
for stellar seismology. In photometry, it is
possible from space
to track modes with amplitudes around 1ppm with
very moderate apertures (~30 cm for CoRoT and for
objects with mv$\sim$6).
Space also enables very high duty cycles and extended runs
(up to 150d with a duty cycle higher than 90\%,
with CoRoT), giving
access to characteristic time scales out of reach
from the ground. The past two decades have seen
an uninterrupted succession of proposals for
national and international space projects
dedicated to stellar seismology.
The whole community is longing for such a unique point of view.
The results from the Canadian experiment
MOST (Matthews 2005) have raised an animated debate, 
illustrating the fact that the field is entering a
new area. In a very close future (launch autumn 
2006), CoRoT (Baglin et al. 2003) will constitute
a major step in
this domain, hopefully followed by other projects, like Siamois at D\^ome C
(see Bouchy et al. 2005, Mosser et al., this volume), Kepler, 
Plato (see Catala et al. this volume),...

\section{CoRoT Seismology Core Programme - the mainlines}

The CoRoT Seismology Core Programme is addressing objects in a wide range of mass, between 
1~M$\odot$ and $\sim$15~M$\odot$. It is highly focussed on main sequence 
evolution stage which represents 90\% of the stellar lifetime.
As commented herebefore, it is time now to make a qualitative step in the 
understanding and description of this stage of evolution. Several theoretical developments
are proposed which need observational constraints. From early studies and as illustrated in examples
given hereafter, it comes out that frequency measurements with precision of the order of 0.1$\mu$Hz, 
are sensitive to
the detailed structure of stars in this evolution stage and susceptible to bring
valuable discriminant tests for it. 
CoRoT seismology observation programme is thus intended to bring 
observational material of this quality for a significant set of objects.
This is at the origin of one of the most specific caracteristics of CoRoT: the possibility to dedicate
long runs (up to 150 days) to the same field. 

Besides this guideline, an interest has been clearly affirmed for complementary shorter runs 
allowing an exploration
of the pulsational behavior at the micromag level across the HR diagramme.
The mission profile has thus been built around (at least) 5 long runs of 150 days each,
completed with approximately the same number of shorter runs ($\sim$20 days).

\subsection{CoRoT and Solar-like pulsators}

     The amplitudes observed in the Sun have been used as a dimensioning  constraint 
to fix CoRoT specifications. This was initially justified by the theoretical prediction 
that amplitudes of solar-like pulsators increase with convective velocities (and thus with
L/M), see e.g. Houdek et al (1999), and this has been
confirmed since then by observations from the ground (see Samadi et al 2005).
Solar like candidates  for CoRoT have thus been selected among F stars on or 
near the Main Sequence. They are representative of masses between 1.1 and 1.5~M$\odot$,
where the structure is changing significantly from small or no convective core
associated with an extended outer convective zone to an important convective core and
a tiny envelope.

\begin{figure}[ht]
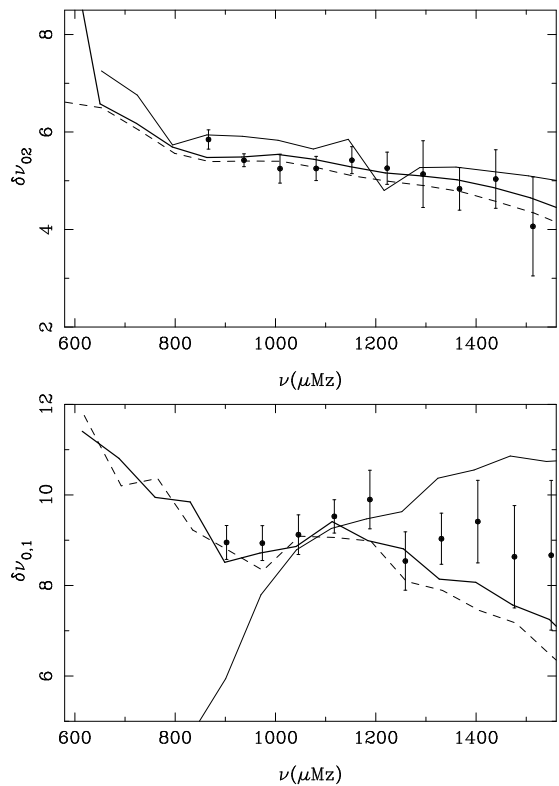

\includegraphics[angle=0,width=0.9\hsize]{II_2_a_sismo_fig3a.ps}
\includegraphics[angle=0,width=0.9\hsize]{II_2_a_sismo_fig3b.ps}
\caption{
Small frequency spacings $ \delta \nu_{02}$
and  $  \delta \nu_{01}$ as a function of the frequency, for two
main sequence models M1 (heavy and dashed lines) and one post main
sequence model (normal line) of HD 45067, compared to simulated data
of HH3 exercise. Error bars result from the analysis of the simulated data.
\label{figJP}
}
\end{figure}

{\bf Hunting for a 0.1~$\mu$Hz precision on frequency measurement.} 
Several early prospective works have shown that eigenfrequencies, if determined with
a precision of the order of 0.1$\mu$Hz were sensitive and discriminant in terms of
physics options in the modelling of these objects. This is the case of Michel et al.
(1995) who investigate the effect of slight variations of overshooting amount, mixing-length,
metallicity, etc...
This forward approach has been investigated further in the framework  
of 'Hare and Hounds' (H\&H) exercises reproducing dimensioning factors of the expected CoRoT
observations.  
Using simulated data of the H\&H3 exercise for instance, Provost et al. (2002,
see also Berthomieu et al. 2003, Provost et al. 2000), 
investigates the potential of classical frequency indexes 
$  \delta \nu_{02}= \nu_{n,\ell=0} -  \nu_{n-1,\ell=2}$
and $  \delta \nu_{01}=2 \nu_{n,\ell=0} - (\nu_{n,\ell=1}
+ \nu_{n-1,\ell=1})$.
As illustrated in Fig.~\ref{figJP}, they concluded that $\delta \nu_{01}$
is a powerfull indicator to discriminate main sequence and post main sequence evolution
stages. More examples can be found in Appourchaux et al. (this volume) and
Berthomieu et al. (this volume). 

\begin{figure}[ht]
   \begin{center}
     \epsfig{file=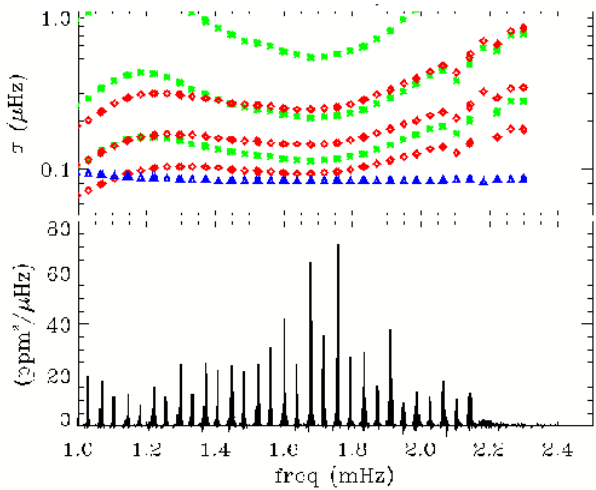, width=8cm}
   \end{center}
\caption{ Lower panel: simulation of the pure sismic signal expected for HD49933.
Upper pannel: Estimates of the 1-$\sigma$ precision on the determination
of eigenfrequencies, for case a, b and c (resp. triangles, rhombs,
stars) as described in the text (from Michel et al. 2006).\label{libb49933}}
\end{figure}

In order to explore what the final precision on frequencies determination could be,
using the simulation tool developed by Baudin \& Samadi (this volume), Michel et al. 
(2006) gave an illustration
(see Fig. ~\ref{libb49933}) 
of what kind of spectra and performances are expected for a solar-like target planned for a long run (150 days). 
Three cases are illustrated: a) the reference case: precision is estimated considering
only photon noise and taking a 1$\mu$Hz generic value for linewidth; b) as case a, but
linewidth are from Houdek et al. (1999); c) same as b, but granulation noise contribution
is considered in addition to photon noise. In order to illustrate the impact of the uncertainty
on linewidth estimates, for case b and c, results are also shown for
twice and half the values of the estimated linewidth.  These results
confirm the fact that the granulation noise as estimated here following Harvey (1985), might 
be a significant factor compared with photon noise. However, the expected precision 
on frequency determination remains of the order of 
a few 10$^{-7}$ Hz, (below 0.7$\mu$Hz in the worst case considered here).

{\bf Tackling specific structural features} 
One of the means to investigate the structure of the stellar interior
from the oscillation frequencies without resorting to explicit modelling
of the star is to utilise the oscillatory signal in the frequencies to
determine the acoustic depth $ \tau_d = \int^R_{r_d} \mathrm{d}r/c$, of
a sharp feature, lying at a radius of $r_d$, $c$ being the sound speed
and $R$ the total radius of the star. This signal first commented by Gough (1990)
 can be amplified by
using the second differences of the frequencies, which can then be fitted
to a suitable oscillatory function to determine the acoustic depth,
$\tau_d$. 
Ballot et al. (2004) have shown that the long runs of CoRoT would allow to
to extract the position of the 
bottom of the convective 
zone (BZC) of solar-like stars,
within an accuracy of around
5\% for the majority of solar-like targets.

\begin{figure}
\begin{center}
\includegraphics[width=0.9\hsize]{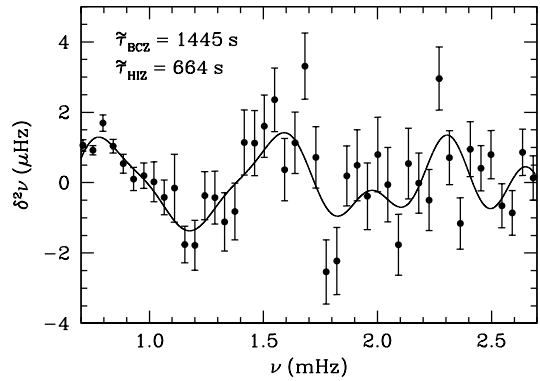}
\end{center}

\caption{
The oscillatory signal in the second
differences of the (simulated) frequencies of HD49933 (shown as data
points with respective errorbars) are fitted by a double-oscillatory
expression ({\it solid curve}) to extract the acoustic depths of
the base of the convective envelope ($\tau_\mathrm{BCZ}$) and the second
helium ionisation zone ($\tau_\mathrm{HIZ}$), (from Mazumdar 2005).
\label{figmaz}
}
\end{figure}

Mazumdar (2005) applied this technique to the simulated CoRoT
data for the primary target star HD49933 to correctly extract the
acoustic depths of the base of the convective envelope and of the second
helium ionisation zone of the input stellar model. Fig.~\ref{figmaz} shows the
functional fit to the simulated data with errors.

Such measurements with CoRoT data would constitute a strong constraint for better understanding 
convection in stars. 

\begin{figure}[ht]
   \begin{center}
     \epsfig{file=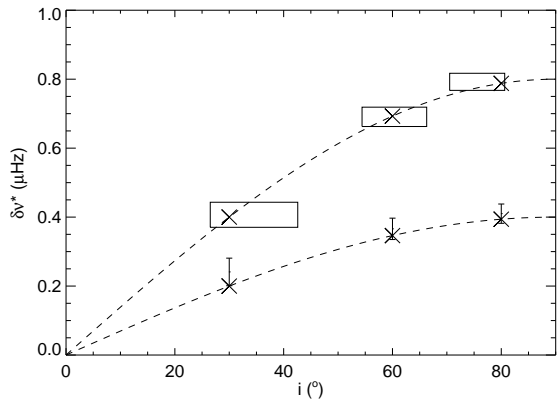, width=8cm}
   \end{center}
\caption{ 
 Biases and error bars for
parameters $i$ and $\delta\nu^{\star}=\delta\nu\sin i$
obtained with a multi-mode fitting
for 6 different simulated configurations 
($\delta\nu_0=\Omega/2\pi= 0.4, 0.8\:\mu$Hz \& $i_0$ = 30, 60 and 80 degr).
The crosses mark the expected values
$(i_0,\delta\nu_0^{\star})$. For $\Omega=2\:\Omega_{\odot}$ cases,
the boxes indicate the mean results and their dispersions. For
$\Omega=\Omega_{\odot}$ cases, only error bars on $\delta\nu^{\star}$
are plotted because of the absence of good determinations of $i$.
The two dashed lines are isorotations
$\delta\nu=\delta\nu_0=0.4$ and $0.8\:\mu$Hz
(from Ballot et al. 2006).\label{JBallot}}
\end{figure}

{\bf Rotation and inclinaison.}
Gizon \& Solanki (2003) have studied the possibility of constraining 
both rotation rates (assumed to be rigid in first approximation) 
and stellar axis inclination from
p-modes of stars spinning as slowly as two times the solar rate. 
Ballot et al. (2006) have investigated the value of using several
modes simultaneously to increase the accuracy, especially at low angle  
(Fig.~\ref{JBallot}),
and pointed out the strong correlation between the estimates of these two parameters.

\begin{figure}
  \resizebox{\hsize}{!}{\includegraphics{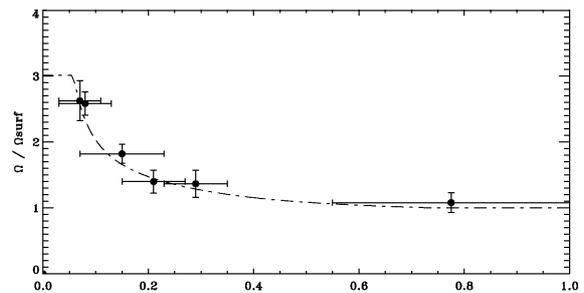}}
  \caption{
Dashed line: input profile for rotation rate (normalized to the surface here). Black dots and associated error
bars: values obtained for inversion at different radii, considering representative noise and 
using rotational kernels computed
with a trial stellar model showing large separation close to the ones of the input model. 
(from Lochard et al. 2005)
}  
\label{JLinvrot}
\end{figure}

Considering rotation profiles, and as stressed by the solar case, rotation inversion possibilities 
toward the center of the stars
are very dependent on the detection of g or mixed modes. From the beginning, the idea 
thus has been to address this problem mostly with classical pulsators known to show this kind of modes. However, Lochard et al. 
(2005) have shown (Fig.~\ref{JLinvrot}) that for appropriate solar-like targets observed with CoRoT, 
such modes being expected, it should be possible
to have an estimate of the gradient in the rotation profile.

\begin{figure}[ht]
   \begin{center}
     \epsfig{file=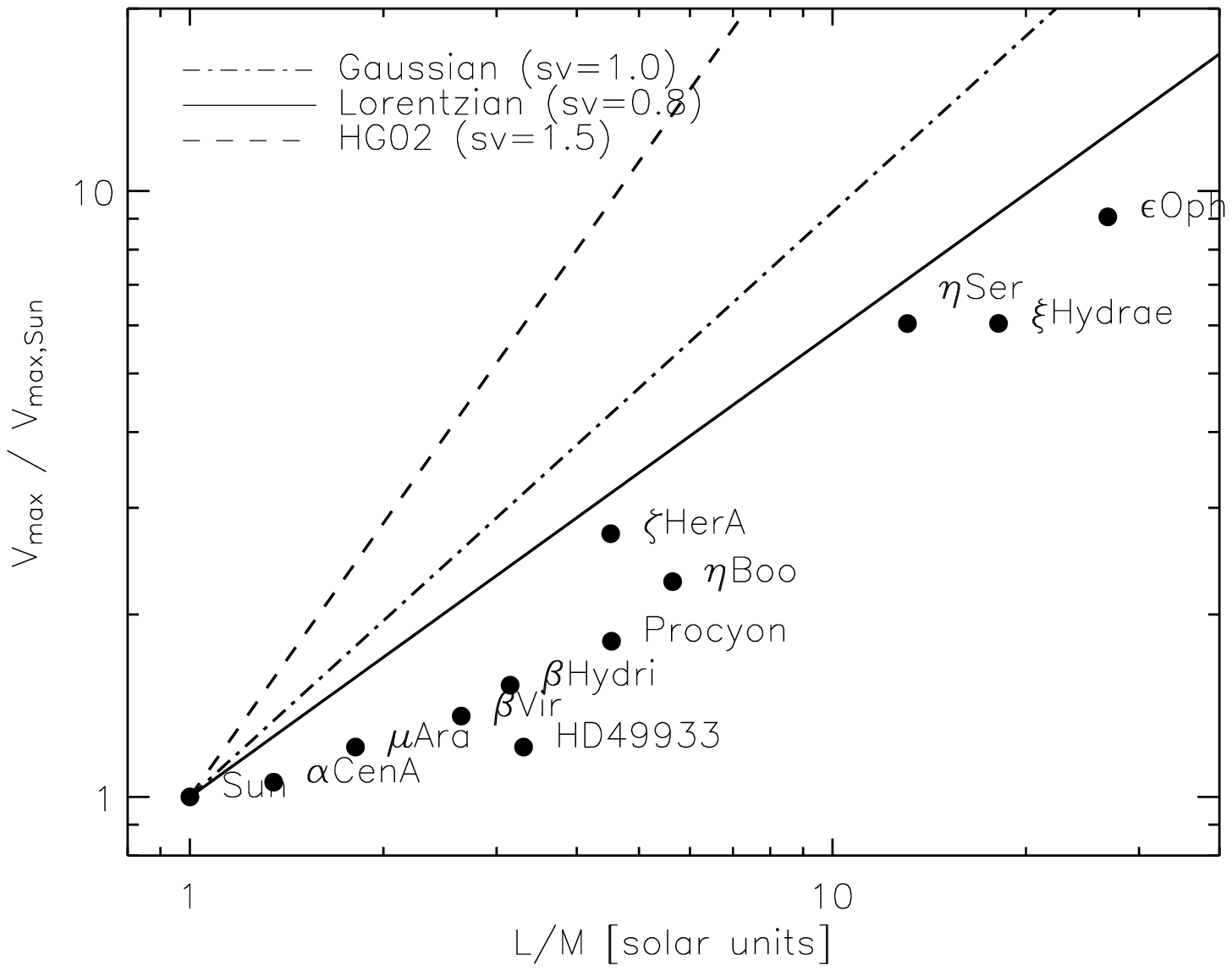, width=8cm}
   \end{center}
\caption{ 
Maximum of the mode amplitudes ($V_{\rm max}$) relative to the observed solar
($V_{{\rm max},\odot}=33.1\,\pm 0.9$\,cm\,s$^{-1}$) versus $L/M$ (see text).
Filled circles correspond to the few stars for which solar-like oscillations  have been
detected in Doppler velocity. The lines correspond to
calculations obtained by Samadi et al. (2005)
 assuming Lorentzian eddy time-correlation functions (solid line)
and Gaussian function (dot dashe line).
For comparison the dashed line shows the result by Houdek \& Gough (2002),
using a Gaussian function and classical Mixing Length Theory.
 \label{RSvmax}}
\end{figure}

{\bf Amplitudes and convective transport.}
As shown by Samadi et al (2005), the study of amplitude distribution in this
domain of the HR diagram can be used  to constrain the properties of stellar turbulent convection.
Indeed, the square of the mode amplitude, $V^2$,  is proportional to ${ \cal P} \, / \, \eta$
where  ${\cal P}$ is the rate at which energy is
supplied by
turbulent convection  and $\eta$ is the rate at which the mode is damped.
Using several 3D simulations of stars,  Samadi et al (2005) have found that the maximum of the
excitation rate, ${\cal P}_{\rm max}$,  scales as $(L/M)^s$ where $L$ and $M$  are the luminosity
and the mass of the star respectively and $s$ is the slope of this power law.
Futhermore the authors have found that the slope $s$ is very sensitive to the way the
convective eddies are time-correlated. Indeed, the slope $s$ is egal to 3.2 when one models
the eddy time-correlation  according to a Gaussian function and to 2.6 when one models it
according to a Lorentzian function. A comparison of their results using damping rates
($\eta$) from Houdek et al (1999), with available observations
is strongly in favor of the Lorentzian description 
(see Fig. ~\ref{RSvmax}). 

{\bf Diffusion processes.}
Several works have investigated the possibility to constrain chemical inhomogeneities induced by diffusion
processes in stars (Th\'eado et al. 2005, Castro and Vauclair 2006), confirming that with the accuracy 
expected for CoRoT, frequencies would constitute sensitive observables. 
In the case of solar-like stars hosting planets, it has been proposed to use seismology with CoRoT
to distinguish between two possible scenari which are currently considered to explain metallicity excess
in stars with planets 
(Bazot and Vauclair 2004, Bazot et al. 2005, see also Soriano et al. in this volume).

\subsection{CoRoT and classical pulsators}

\begin{figure}
  \resizebox{\hsize}{!}{\includegraphics{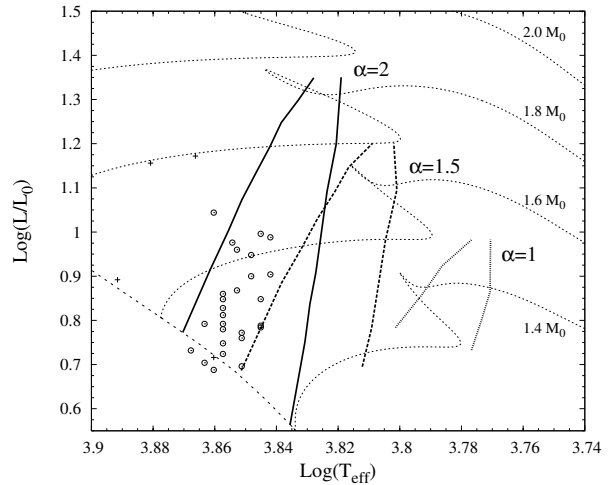}}
  \caption{$\gamma$ Dor theoretical IS for $\ell=1$ modes,
  for three families of models with different values of $\alpha$: 1, 1.5
and 2 obtained
  with TDC treatment (thick lines).
  The small circles correspond to observations of 27 {\sl bona fide}
$\gamma$ Dor stars.
  $\alpha=2$ theoretical IS best agree with observations.}
  \label{gadinsta}
\end{figure}

{\bf Modal stability and convection description.} 
The delta Scuti, beta Cephei, gamma Doradus and PMS stars to be observed
by COROT are auto-driven pulsators. The non-adiabatic modelling of stellar 
pulsation enables us to determine which modes are stable or overstable
and to localize the driving and damping regions inside the stars. In
delta Scuti and
gamma Dor stars, the description of convection and its time-dependent
interaction with oscillations plays a major role in this driving; and in
the beta Cephei stars, it is the metallic content and its location
(where it accumulates due to transport
mechanisms). Hence, the comparison with the observed excited modes
enables us constraining these aspects. A specific analysis can be
performed for each star, comparing the theoretical and observed range of
overstable modes. But also, a general study can be made for each
type of star, comparing the theoretical and observed instability strips
in the HR diagram. As an illustration, we give in Fig. ~\ref{gadinsta} the
theoretical instability strips
obtained for the gravity modes of gamma Doradus stars with the
time-dependent
convection (TDC) treatment of Grigahcene et al. (2005), for different
values of the mixing-length parameter
alpha (related to the size of the convective envelope) (Dupret et al.
2005). This illustrates how the
description of convection can be constrained by a stability analysis. In
this case, the best agreement is
obtained for models with $\alpha=2$.

{\bf Rotational profile inversion.}
As already mentioned, one of the assets of intermediate and high mass pulsators on the main sequence
resides in the fact that they are expected to show mixed modes. As shown by Goupil et al. (1996),
for such objects, it is possible to build rotational kernels sensitive down to the very
central regions and thus study the angular momentum transfer, one of the key aspect of this evolution
stage for these objects.

{\bf Hot objects} 
Beta cephei stars are main sequence stars with masses around 10 M$_{\odot}$.
Their structure is rather simple: a large convective core surrounded
by a radiative envelope. Their metallicity is close to solar. Their
oscillations are excited by the kappa mechanism, and they exhibit low degree
and low order p and g modes. Their spectrum of oscillations is rather
sparse and they are slow rotators, so that the multiplets due to rotational
splitting can be identified.
It has been shown from ground-based observations that asteroseismology of these
stars can provide precise information on their global parameters,
such as their mass, radius, age, metallicity, overshooting parameter, but also
on their internal rotational law (Aerts et al. 2003).
However, several problems remain unsolved. Indeed, some of the observed modes
of oscillation cannot be excited using standard stellar models
(Ausseloos 2004); they also
present variable surface enhancements of nitrogen, which are hard to explain
given that they are slow rotators (Morel et al. 2006).
It is necessary to include non-standard physics to explain these observations.

\begin{figure}
  \resizebox{\hsize}{!}{\includegraphics{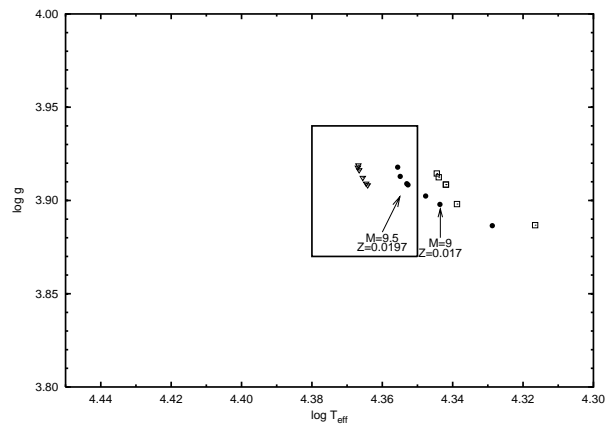}}
 \caption{Positions of the stellar models which fit exactly two observed
frequencies of oscillation
of the $\beta$ Cephei star HD~129929 in a $\log T_{\rm eff}$ - $\log g$
diagram.
The squares, dots and triangles are obtained for models with
$\alpha_{\rm ov}=0, 0.1, 0.2$, respectively. The observational error box
from photometry is also given.} 
  \label{AT129929}
\end{figure}

A hare-and-hound exercise on beta cephei stars was done to prepare the COROT
mission (Thoul et al 2003). The conclusion reached was that due to the
simplicity of these stars
it was possible to reconstruct the original star to a very high level of
precision (Fig.~\ref{AT129929}). It was also concluded that in order to discriminate between
different models, it was useful to observe more modes, including modes of
degree $\ell >$2, which are not
observable from the ground. In addition, several multiplets have to be
observed in order to probe different depths in the star if we want to get
information about the internal rotation law.

In order to prepare the analysis of the data that COROT will provide on
beta cephei stars, a database is being constructed, which contains stellar
models and their oscillation frequencies. This database is described in details
Thirion and Thoul (this volume).
 
{\bf g-modes pulsators.}
$\gamma$ Doradus stars pulsate in the asymptotic g-mode regime. This
makes it possible to obtain relevant physical information through the
analytical expressions that the asymptotic theory provides. In this
particular case, the eigenfrequency is given by (Tassoul 1980)

\begin{equation}
\sigma_{asymp}={\sqrt{\ell(\ell+1)}\over \pi(n+1/2)}\int_{r_a}^{r_b}
{N\over r}dr
\end{equation}

\noindent where $n$ is the radial order, $\ell$ the spherical order,
$r_a$ and $r_b$ are the lower and upper limits of the radiative
envelope of this stars and $N$ is the Brunt-V\"ais\"al\"a frequency.

Therefore, as suggested by Moya et al. (2006),
the ratio between two frequencies in the asymptotic regime
depends only on the radial and spherical orders, 
taking the form

\begin{equation}
{\sigma_1\over\sigma_2}={n_2+1/2\over n_1+1/2}
\end{equation}

This allows us to estimate the radial order of observed $\gamma$
Doradus frequencies. With at least three observed frequencies we can
infer, through this procedure, some possible values of the radial
order of each frequency. Once an estimate is fixed, the asymptotic
expression provides a value for $I$ the Brunt-V\"ais\"al\"a integral in (1). This
gives us a new observable to be fitted by models.

\begin{figure}[ht]
   \begin{center}
     \epsfig{file=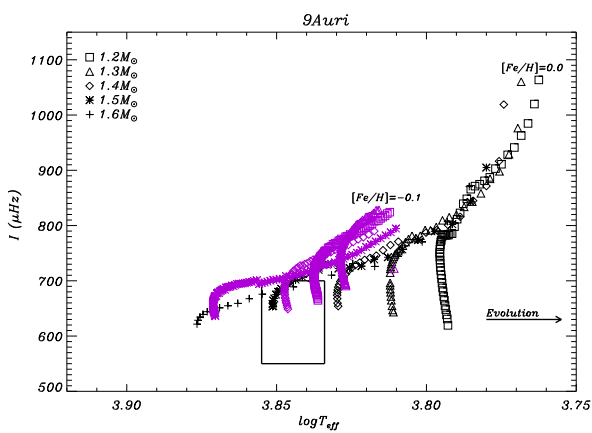, width=8cm}
   \end{center}
\caption{
$I$-Teff diagramme (cf text). The box represent the estimated value of the integral from
observations, and the dots are the model predictions for different
masses, metallicities, overshootings and evolutionary stages. 
From Moya et al. 2006. 
\label{GamDor}}
\end{figure}

If we display, for a given observed star, the estimated integrals as a
function of the effective temperature, a figure giving a new
constraint for the modeling of these stars is obtained (see Fig.~\ref{GamDor} 
for an example in the particular case of 9
Aurigae). 
This
technique has been successfully used for different $\gamma$ Doradus 

\begin{figure}[ht]
   \begin{center}
    \includegraphics[angle=90,width=0.9\hsize]{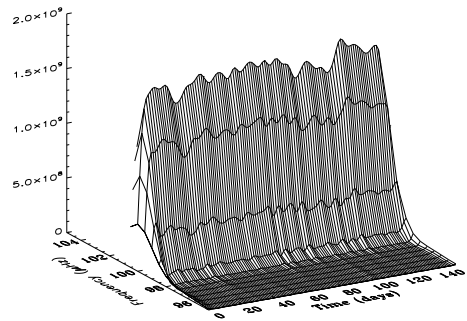}
   \end{center}
\caption{ 
Time/frequency analysis of a 150 day
simulation of a sinusoidal oscillation of constant amplitude (400
ppm) at the frequency $\nu=100$~$\mu Hz$. 
Photon noise,
granulation and activity signals have been computed using the simuLC
tool described in Baudin et al. (this volume)
 \label{FBTF2}}
\end{figure}

{\bf Time/frequency analysis.} 
For several class of classical pulsators, the question of the variability of the mode amplitudes
is still an open question lacking seriously well suited observations.
CoRoT long runs will offer a unique opportunity to apply time/frequency analysis
and address this question.
In order to get an estimate of the precision that one can
expect for amplitude variations determination, F. Baudin made the simulation
illustrated in  
Fig.~\ref{FBTF2}: The application of 
time/frequency analysis to a 150 day
long simulation of a sinusoidal oscillation of constant amplitude 
(400 ppm, i.e. representative of a lowest limit of what is seen from the ground) 
at the frequency $\nu=100$~$\mu Hz$. 
In this simulation, a m$_V$=8 star is considered. 
The apparent power variations are due to the presence of noise
(including activity and granulation). Their standard deviation
corresponds to a 3\% variation in power, for a time resolution of 5
days. Of course, this precision on the power variations will vary with
the choosen time resolution of the analysis.

\subsection{CoRoT sample:}

Among the criteria used to select targets, observational performances have of course been considered.
In this respect, a rapid estimate shows that for classical pulsators, in the hypothesis of stable frequencies
and amplitudes over 5 months, a gain by a factor ~500 at least can be made down to m$_v$=9 in terms of S/N, compared with
what is currently achieved from the ground (see e.g. Michel et al. 2006). This lets room to choose these targets 
according to other criteria,
like evolution stage, rotation rate, binarity,... 

In the case of solar-like pulsators, the situation is less 'comfortable' and apparent magnitude
always appears as a high priority parameter for selection. This is why these objects appeared very
early and with a strong priority in the process of fields selection. 
First, the so-called Principal candidate stars have been selected with the conservative
criterion based on observed solar amplitudes. Then, three criteria have been defined
(Samadi et al. 2004)
to evaluate the interest of secondary candidates to be selected in the field around.
These criteria are intended to determine for which objects, a 'significant' number of modes
can be expected to be measured with a given minimal precision.
The amplitudes are estimated 
following Samadi et al. (2005) for different values of the linewidths, as
commented hereafter.
The noise level is obtained considering photon noise for CoRoT. A 150d duration of the run
is assumed. Then, following Libbrecht (1992), an estimate of the frequency precision that could be obtained
for a peak with half the maximum expected amplitude is derived.
\begin{figure}[ht]
   \begin{center}
     \epsfig{file=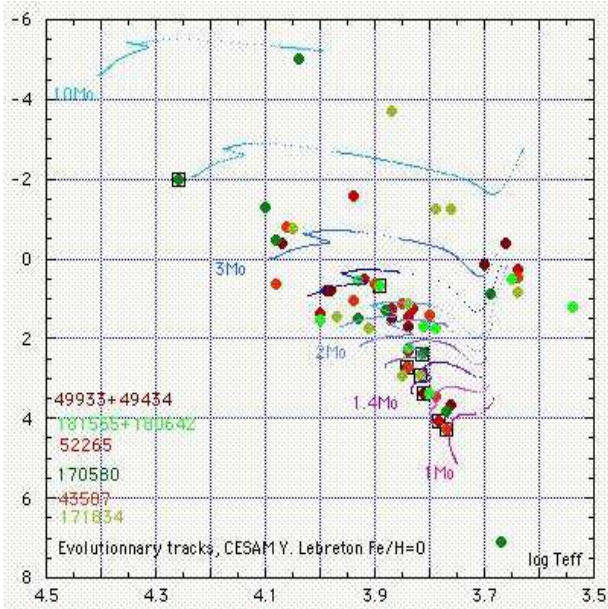, width=8cm}
   \end{center}
\caption{ HR diagramme built with an illustrative selection of candidates potentially observed
during 150 days with CoRoT. From Michel et Baglin (2005).
 \label{corothr}}
\end{figure}

\begin{figure}[!ht]
   \begin{center}
     \epsfig{file=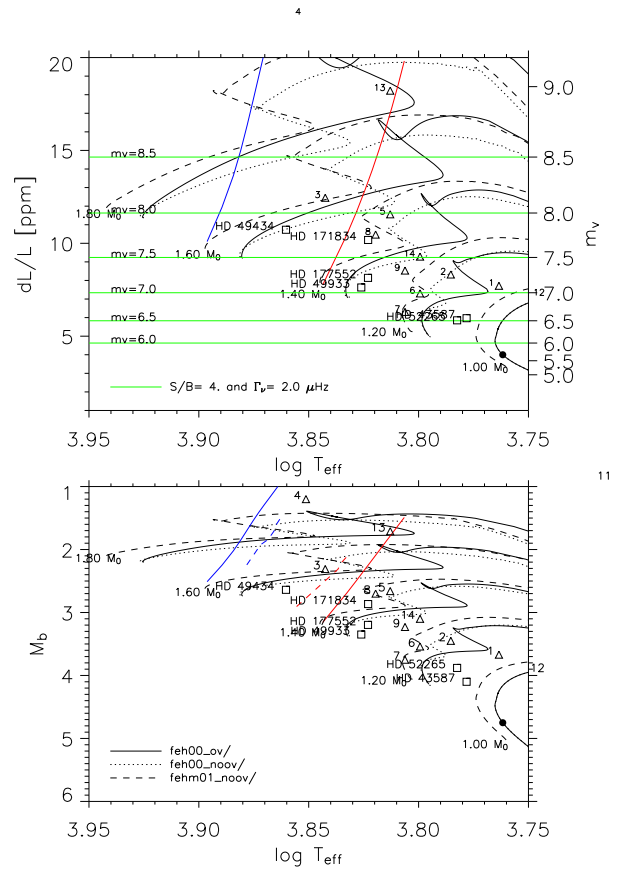, width=8cm}
   \end{center}
\caption{ criterion 1. Candidate targets are attributed a mass and a luminosity by considering their location
in the HR diagramme (lower panel) compared with evolution tracks. In the upper panel, the same evolution tracks 
are put in a diagramme  with maximum amplitude estimates
in ordinates (left axis). On the right axis are given values of observational magnitude m$_V$. Each of them (e.g.
7.0) separates the diagramme in two parts. If a candidate target is found in the upper part (above the 
corresponding green line), and if its observational magnitude m$_V$ is lower than the limit value associated 
with this line (here 7.0), then the target satisfy the criterion. For instance candidate 1 would satisfy the present
criterion if it is brighter than m$_V$=7.0. 
\label{RSsolselec}}
\end{figure}

{\bf Criterion 1} is our standard. It is refeering to objects for which, assuming a generic
2~$\mu$Hz linewidth ($\sim$1.8 days lifetime),
modes with amplitudes higher than half the expected maximum (i.e. a significant
amount of peaks) would have frequency measured with precision better than 0.25~$\mu$Hz (associated with a
signal to noise ratio S/N=4, as defined by Libbrecht (1992)). 
The application of this test is illustrated on Fig.~\ref{RSsolselec},

{\bf Criterion 2} is an extrapolation of criterion 1 in the pessimistic eventuality of a 5~$\mu$Hz linewidth
($\sim$0.7 days lifetime), 
For objects satisfying this test, modes with amplitudes higher than half the expected maximum (i.e. a significant
amount of peaks) would still have frequency measured with precision better than 0.4~$\mu$Hz (associated with a
signal to noise ratio S/N=4, as defined by Libbrecht (1992)). 

{\bf Criterion 0}, the lowest one, also assumes a 2~$\mu$Hz linewidth, but corresponds to a  S/N =1 value, which 
would allow the detection of a significant number of oscillation peaks, but is not expected to bring very precise
frequency values. 
Roughly all stars brighter than m$_V$=7.5-8 satisfy this
criterion.

Michel and Baglin (2005), with a preliminary selection of potential targets  gave a flavor of how the 
CoRoT sample of stars observed during long runs could distribute in an HR diagramme.
As shown in Fig.~\ref{corothr}, it is possible to obtain a reasonable scan of the domain of
interest in the HR diagramme. An updated picture of the target selection state of the art can be found
in Michel et al. (this volume).

%\clearpage

%\begin{acknowledgements}

%text

%\end{acknowledgements}

\clearpage

\end{document}